# Filovirus Glycoprotein Sequence, Structure and Virulence

J. C. Phillips

Dept. of Physics and Astronomy, Rutgers University, Piscataway, N. J., 08854

Abstract

Leading Ebola subtypes exhibit a wide mortality range, here explained at the molecular level by using fractal hydropathic scaling of amino acid sequences based on protein self-organized criticality. Specific hydrophobic features in the hydrophilic mucin-like domain suffice to account for the wide mortality range.

**Significance statement** Ebola virus is spreading rapidly in Africa. The connection between protein amino acid sequence and mortality is identified here.

**Introduction** Ebola filoviruses exhibit ~ 70% glycoprotein (GP) subtype similarities, with wide variations in virulence (from nearly 90% to nearly 0% mortality) [1]. Although the Marburg species has only 30% GP similarity to Ebola, its length and viral morphology are similar, and it also has high mortality levels ~ 50% [2,3]. The static GP domain structure of the most virulent Ebola subtype, Zaire or ZEBOV, whose structure bound to an antibody is shown in Fig. 1 of [4], is the basis of many studies [5-9]. The molecular basis for explaining the widely differential pathogenicity of the filoviruses, which depends on many steps from membrane penetration to selective disruption of cell-cell binding, remains elusive [1,6].

Two GP mechanisms have been suggested as possible origins of virulence variations, different folds (Ebola Zaire is the only structure known) [2], and different flexibility [6]. Ebola subtypes share 70% sequence similarity, and subtypes with more than 40% similarity generally have similar folds [10]. A characteristic feature of filoviruses is a large, disordered, and highly glycosylated mucin-like domain (MLD), which is a natural candidate for disruption of cell-cell binding [6]. Here we present the results of fractal hydropathic analysis which quantifies the subtype dependence of Ebola MLD flexibility on length scales smaller than the MLD length (~ 150 aa). Ebola and Marburg sequence similarity is only 30%, but our analysis shows how the functional similarities arise, and reveals quantitative differences in both their MLD and their receptor domains (RD).

The surface interactions that determine cellular variations in filovirus virulence are monitored at the molecular level by fractal hydropathic analysis [11-13], which is well-suited to treating both short-range [13] and long-range elastic water-protein interactions [14]. The fractal aa-specific hydropathicity ψ(aa) from the MZ scale [11] is averaged over a sliding window length W to describe elastic interactions with the range W/2. ψ(aa,W) has described the evolution of lysozyme *c* with W = 69 [14] and flu GP virulence [15]. Ebola GP has 676 aa, and its MLD has been defined as 313-464 [4,6]. The choice W = 115, with range 57, shows large differences between the major subtypes of Ebola as well as Marburg, and a similarly long range was effective for flu hemagglutinin [15].

**Results and Discussion** The overall GP profiles of three subtypes with variable virulences (ZEBOV, 80-90% mortality, SEBOV, 40-60% mortality, and REBOV, ~ 0% mortality) are shown in Fig. 1. With the MZ scale, hydroneutral is near 155, so the ZEBOV profile consists of two hydroneutral peaks separated by a deep hydrophilic minimum associated with the MLD. The disordered structure of the MLD is highly variable (~ 30% similarity using BLAST) and may be responsible for most of the widely differential pathogenicities. The most virulent subtype ZEBOV shows a deep, nearly smooth "V" minimum near 420, while the least virulent REBOV subtype shows a similar minimum with a wide secondary hydrophobic peak (SHP) centered near 343. The intermediate SEBOV subtype has a similar secondary hydrophobic peak centered near 437. These secondary hydrophobic peaks stabilize the soft hydrophilic ZEBOV V hinge, which becomes a strong candidate for explaining quantitatively the very strong ZEBOV virulence. One can also suppose that the unencumbered ZEBOV deep, nearly smooth "V" minimum near 420 enables the ZEBOV GP to function as a hydroneutral-hydrophilic wedge that pries apart cell-cell interfaces.

The REBOV secondary hydrophobic peak is near the C end of the 227-313 glycan cap; this peak could rigidify the cap. The well-preserved (~ 70%) glycan cap itself is connected to the receptor binding plateau centered near 150, and may play an essential part in receptor binding. The SEBOV secondary hydrophobic peak is centered near the functionally less critical ZEBOV minimum, and it is plausible that it should reduce mortality but not eliminate it.

It has been suggested that GP-mediated steric shielding of host cell surface proteins is an important contributing factor in filovirus cell unbinding [6]. MLD N-linked glycosylation sites are listed by Uniprot, as calculated from the Asn-X-(Thr Ser)Y [X,Y ≠ Pro] rule [16]. These MLD sites are highly variable, as shown for three Ebola subtypes in Fig. 2. As noted there, the most virulent subtype, ZEBOV, has the largest number of sites, but SEBOV (intermediate virulence) has the smallest number. This means that GP-mediated steric shielding is probably less significant for determining virulence than variations in flexibility associated with the MLD SHP (Fig. 1).

In Fig. 1 there is an important ZEBOV shoulder centered on the internal fusion loop (IFL, 511-554), which is enlarged in Fig. 3. EBOV GP is thought to assemble as a metastable trimer of heterodimers on the viral surface [6], before fusing the viral and host cell membranes. The absence of a SHP in the MLD of ZEBOV could destabilize oligomer assembly. The ZEBOV hydrophobic shoulder shown in Fig. 3 could compensate for its extremely flexible MLD. . Alanine mutations at L529 and I544 and particularly the L529 I544 double mutation reduce the hydrophobic shoulder by about 30% relative to the ψ(W) IFL difference between ZEBOV and the less virulent subtypes. Experiment showed that these mutations compromised viral entry and fusion. The nuclear magnetic difference resonance (NMR) structures of the I544A and L529A I544A mutants in lipid environments showed significant disruption of a three-residue scaffold that is required for the formation of a consolidated fusogenic hydrophobic surface at the tip of the IFL [9].

The IFL profile can be used to illustrate the advantage of using the more accurate fractal 2007 MZ hydropathic scale, appropriate to second-order thermodynamics of self-organized criticality (SOC) [12-14], compared to the first-order (water-air) 1982 KD scale [17]. The two scales are similar (R = 0.85), but the 2007 MZ scale is generally more accurate. The KD scale exhibits a similar hydrophobic bulge for ZEBOV subtype in the IFL range 511-554, but it also shows a large difference between the three subtypes above 554, where the heptad repeats occur [4]. In that range the MZ profiles have nearly coalesced, which is more plausible. Note that these profiles are calculated with W = 111, implying elastic interactions over a range of 55 aa.

Experimental data on the Marburg virus are less extensive than for Ebola virus, but for completeness the Marburg and ZEBOV ψ(115) profiles are compared in Fig. 5. Similarities of the V-shaped mucin-like regions are apparent.

**Conclusions** The ability of hydroanalysis to explain the virulence differences between Ebola subtypes may seem surprising, but it becomes less so when one considers the significance of the fractal MZ scale [11,12] as a tool for analyzing interfacial interactions. These can occur both on a large scale at cell-cell interfaces (where the flexibility of the MLD is critical), or on a molecular scale during pre-fusion GP oligomeric assembly, where stability depends on balancing GP hydropathic interactions. The scales of these two interactions differ by many orders of magnitude, but fractal scales can produce scale-free results over many decades, for instance, in amyloidosis [18].

**Methods** The calculations described here are very simple, given the scaling tables of [11] and [12]. They are most easily done on a spread sheet, such as an EXCEL macro. The one used in this paper was built by Niels Voohoeve and refined by Douglass C. Allan.

## References

1. Feldmann H, Geisbert TW Ebola haemorrhagic fever. Lancet **377**, 849-862 (2011).
2. Manicassamy B, Wang J, Rumschlag E, et al. Characterization of Marburg virus glycoprotein in viral entry. Virol. **358**, 79-88 (2007).
3. Timen A, Koopmans MPG, Vossen ACTM, et al. Response to Imported Case of Marburg Hemorrhagic Fever, the Netherlands. Emerg. Infect. Dis. **15**, 1171-1175 (2009).
4. Lee JE, Fusco ML, Hessell AJ, et al. Structure of the Ebola virus glycoprotein bound to an antibody from a human survivor. Nature **454**, 177-182 (2008).

5. Dube D, Brecher MB, Delos SE, et al. The Primed Ebolavirus Glycoprotein (19-Kilodalton GP(1,2)): Sequence and Residues Critical for Host Cell Binding. J. Virol. **83**, 2883-2891 (2009).
6. Noyor O, Matsuno K, Kajihara M, et al. Differential potential for envelope glycoprotein-mediated steric shielding of host cell surface proteins among filoviruses. Virol. **446**, 152-161 (2013).
7. Agopian A, Castano S Structure and orientation study of Ebola fusion peptide inserted in lipid membrane models. Biochim. Biophys. Acta-Biomemb. **1838**, 117-126 Part: B (2014).
8. Lennemann NJ, Rhein BA, Ndungo, E, et al. Comprehensive functional analysis of N-linked glycans on Ebola virus GP1. MBIO **5**, e00862-13 (2014).
9. Gregory SM, Larsson P, Nelson EA, et al. Ebolavirus entry requires a compact hydrophobic fist at the tip of the fusion loop. J. Virol. **88**, 6636-6649 (2014).
10. Krissinel E On the relationship between sequence and structure similarities in proteomics. Bioinform. **23**, 717-723 (2007).
11. Moret MA, Zebende GF Amino acid hydrophobicity and accessible surface area. Phys. Rev. E **75**, 011920 (2007).
12. Phillips JC. Scaling and self-organized criticality in proteins: Lysozyme *c*. Phys. Rev. E **80**, 051916 (2009).
13. Phillips JC Self-organized criticality and color vision: a guide to water-protein landscape evolution. Phys. A-Stat. Mech. Appl. **392**, 468-473 (2012).
14. Phillips JC Fractals and self-organized criticality in proteins. Phys. A **415**, 440-448 (2014).
15. Phillips JC Fractal evolution of hemagglutinin and neuraminidase viral dynamics: A(H1N1 and H3N2) (unpublished, 2014).
16. Gavel Y, Vonheijne G Sequence differences between glycosylated and non glycosylated Asn-X-Thr Ser acceptor sites I implications for protein engineering. Prot. Engineer. **3**, 433-442 (1990).
17. Kyte J, Doolittle RF A simple method for displaying the hydropathic character of a protein**.** J. Mol. Biol**. 157,** 105-132 (1982).

18. Phillips JC Thermodynamic Description of Beta Amyloid Formation Using the Fractal Hydrophobicity Scale.

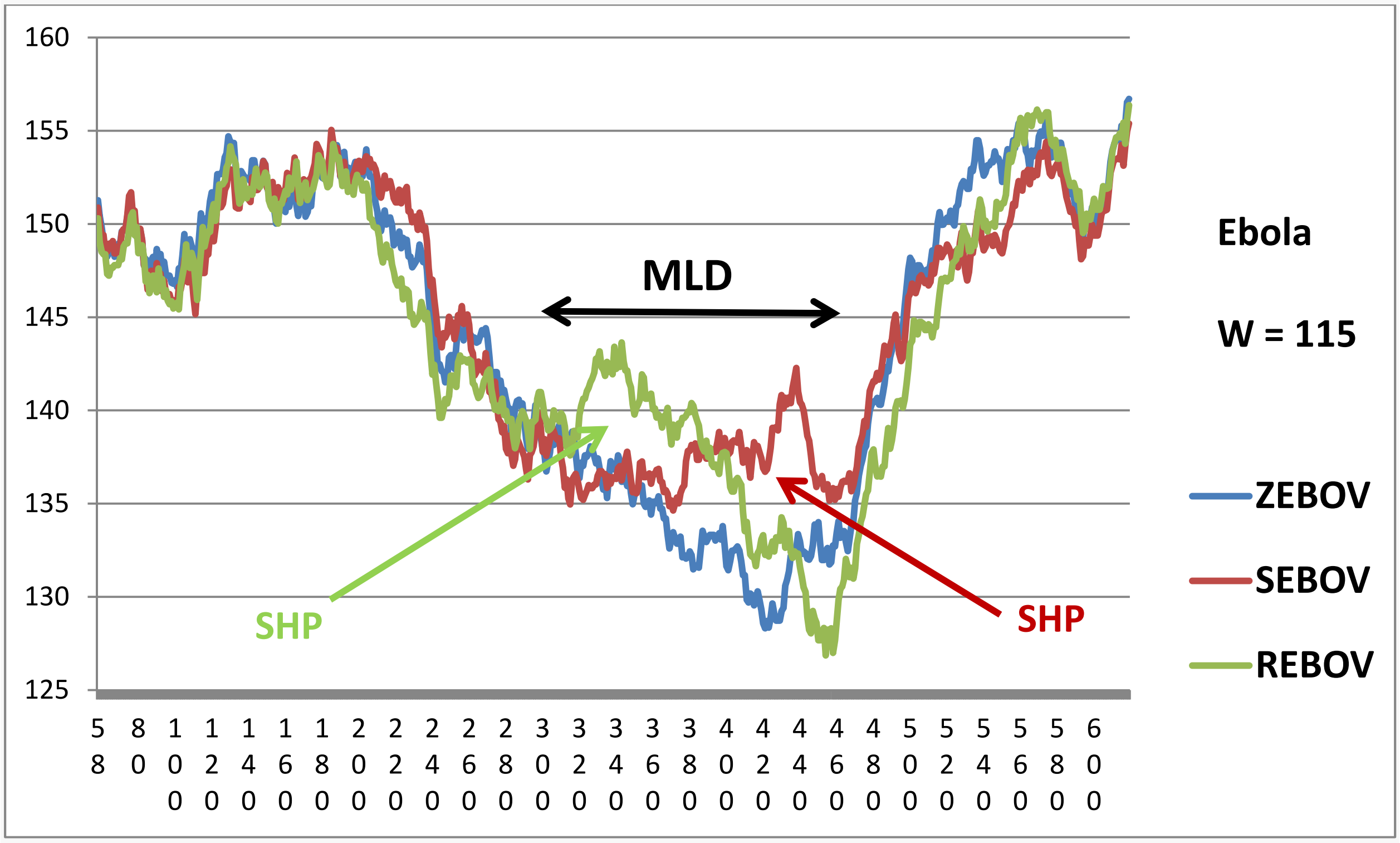

Fig. 1. Profiles of ψ(aa,115) for three Ebola subtypes. The MLD lies in the central hydrophilic valley, between the receptor domain plateau centered on 160, and the transmembrane anchor near 570. Secondary hydrophobic peaks for SEBOV and REBOV subtypes are marked.

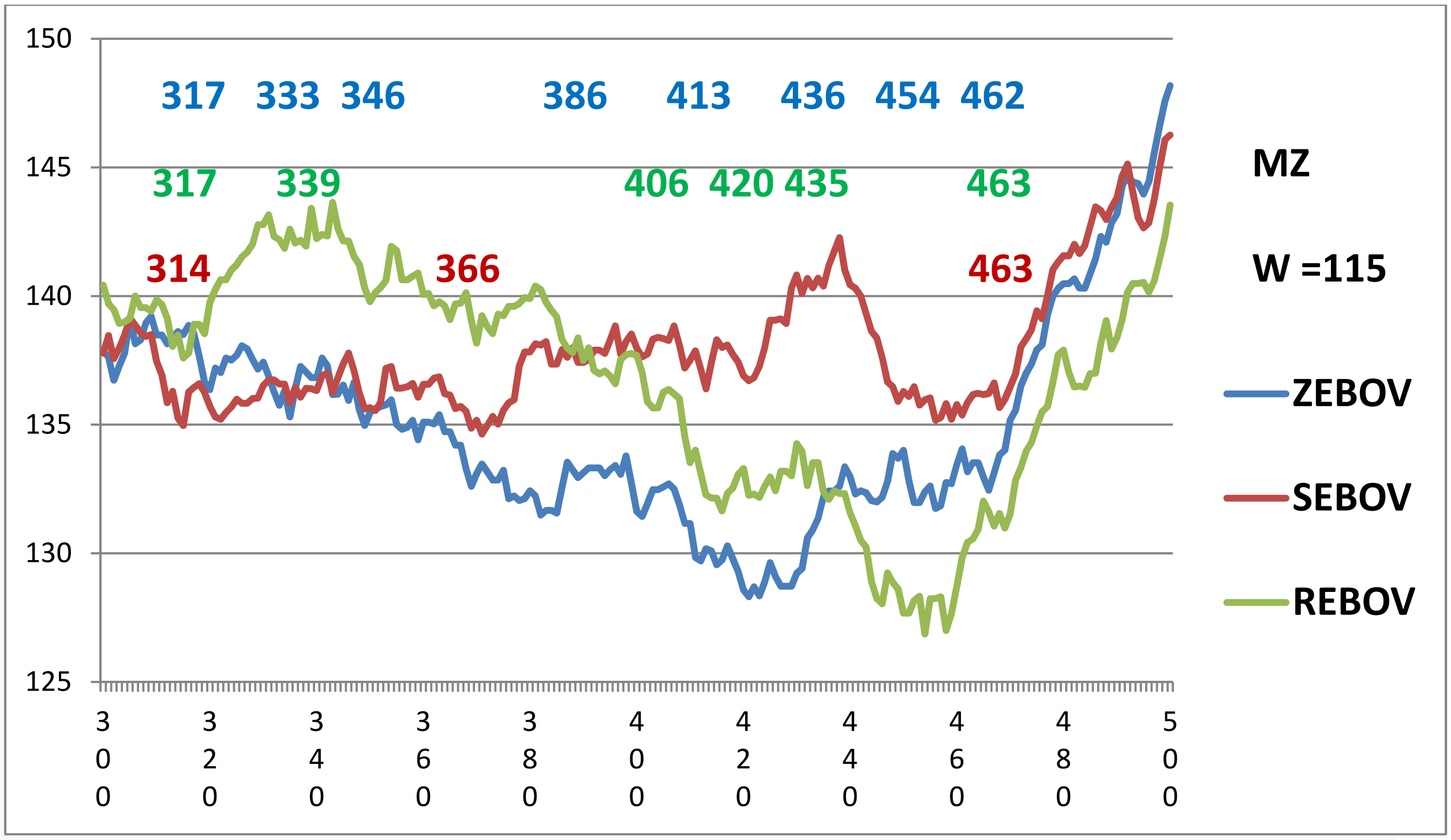

Fig. 2. MLD N-linked glycosylation sites are shown for Ebola subtypes, together with profiles of ψ(aa,115). Their concentration is highest near the deep hydrophilic hinge near 420-460. The most virulent subtype, ZEBOV, has the largest number of sites, but SEBOV (intermediate virulence) has the smallest number.

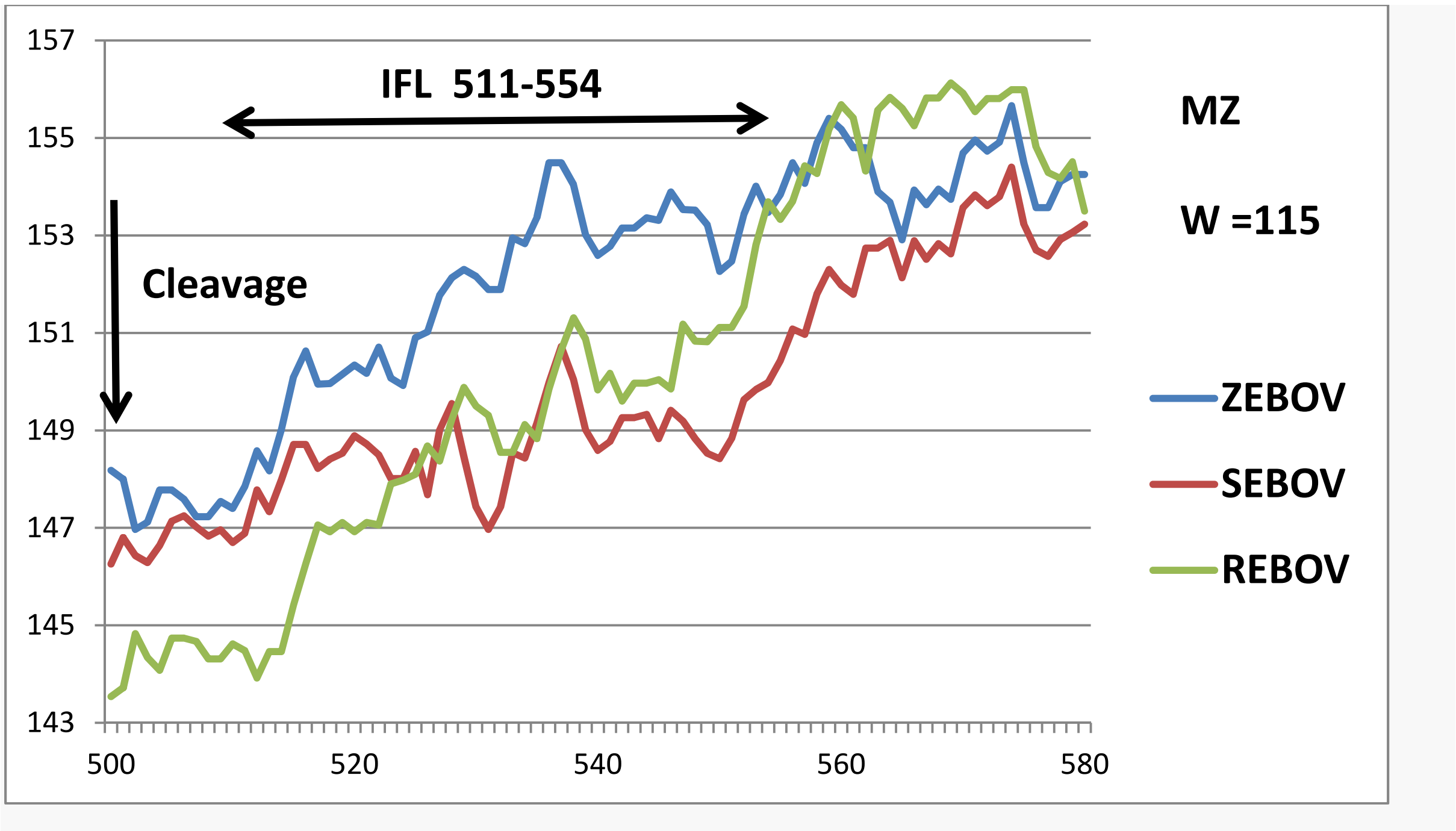

Fig. 3. An enlargement of Fig. 1 in the region of the internal fusion loop (IFL). The excess ZEBOV hydrophobic shoulder is associated with stabilization of oligomer formation [9].

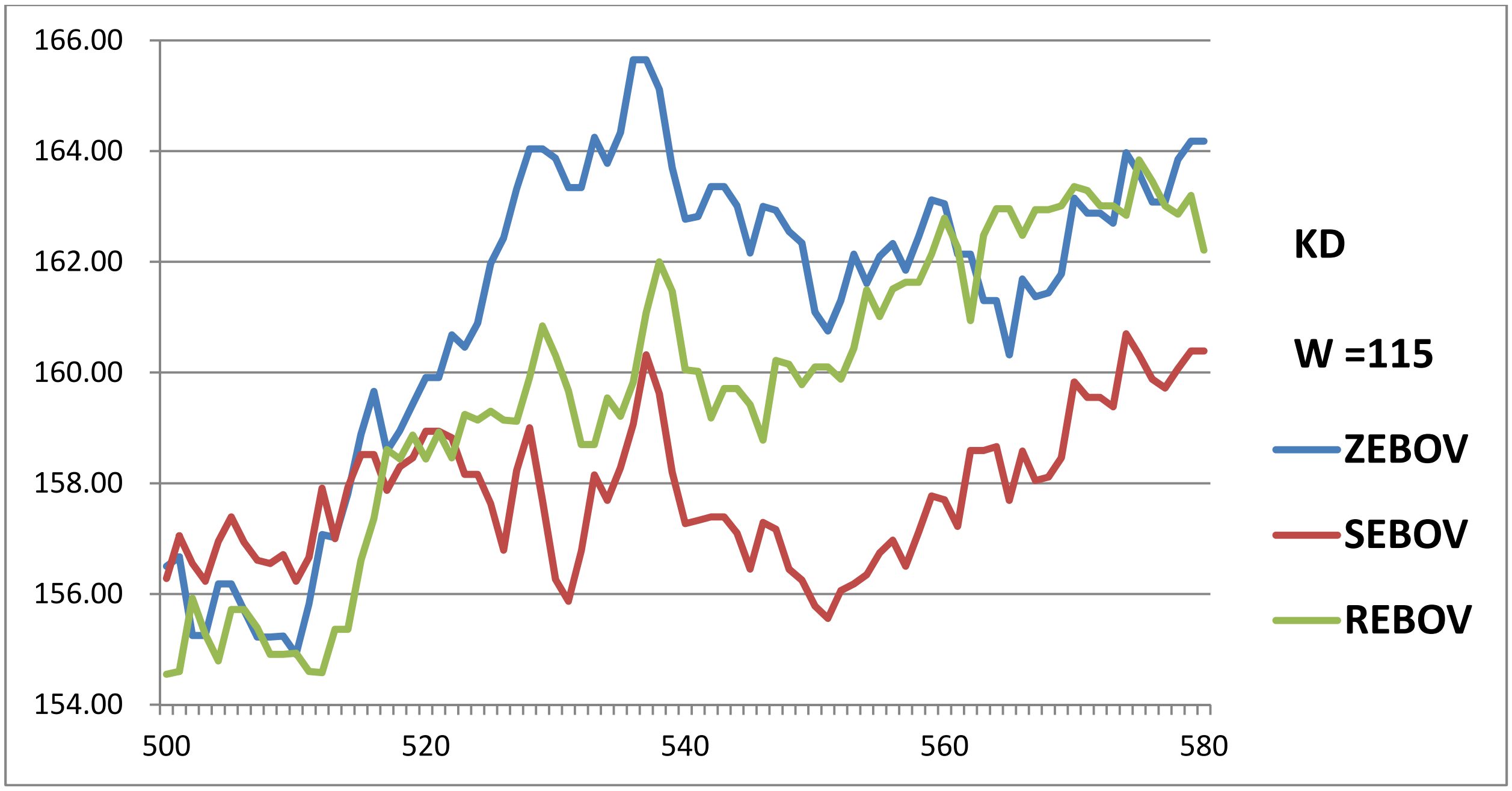

Fig. 4. The IFL profile using the first-order KD scale. Compared to the fractal MZ scale, Fig. 3, there are several similarities, notably the increased hydrophobicity of the ZEBOV subtype in the IFL range 511-554.

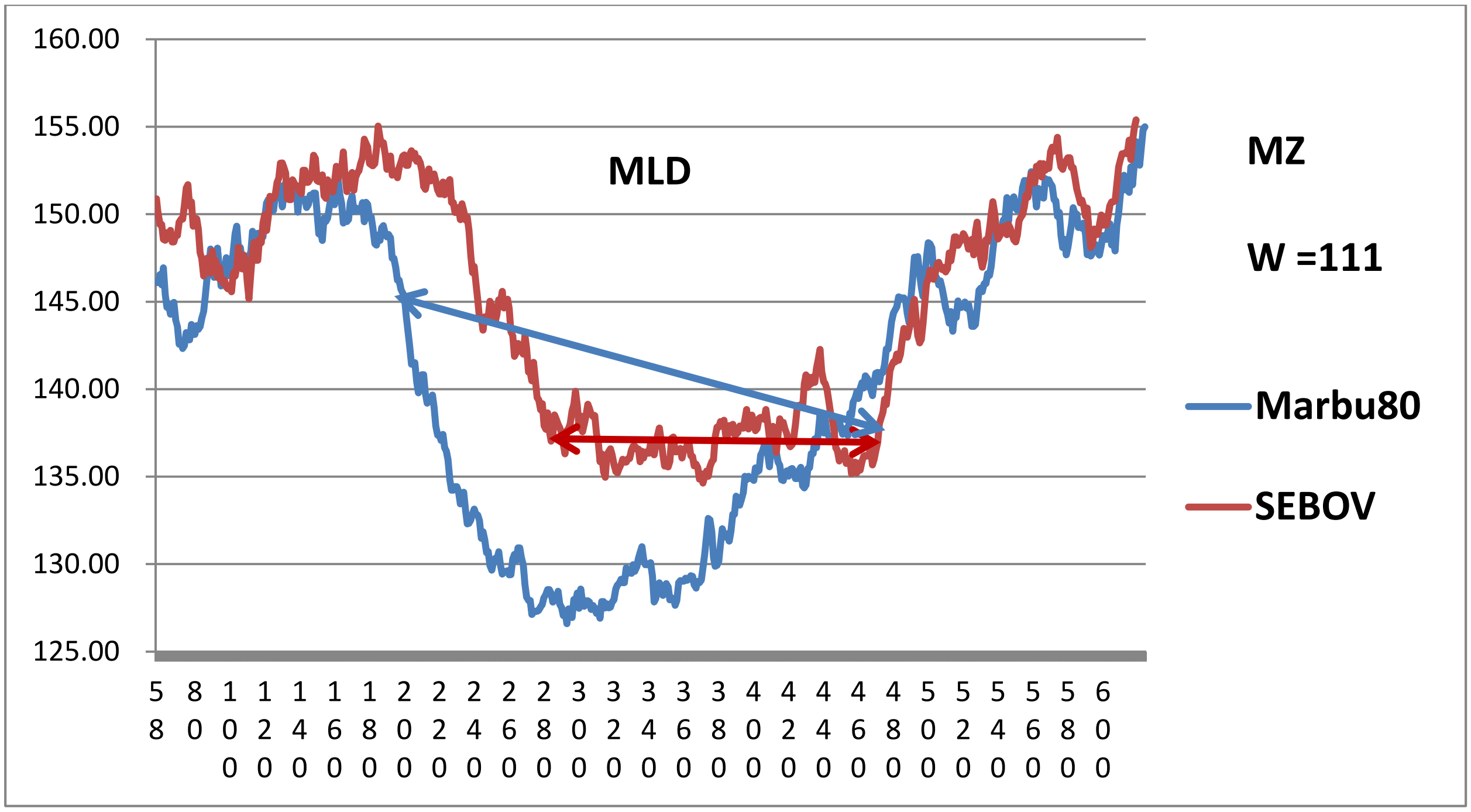

Fig. 5. . Profiles of ψ(aa,115) for SEBOV and Marburg 80, which have similar mortalities near 50%. The MLD domain of Marburg 80 fills the range of high Thr and Ser density, and is seen to be wider than the SEBOV MLD. However, the latter is flatter and more hydrophobic, which may compensate the much larger number ( ~ 15) of N-linked glycosylation sites found in Marburg 80 MLD than in SEBOV (3). Because the former is much more hydrophilic, its sites may be much less occupied.